\documentclass{npw}
\usepackage{graphicx}
\usepackage{mathrsfs}
\usepackage{amsmath}

\newcommand{\lb}{\left(}
\newcommand{\rb}{\right)}
\newcommand{\ls}{\left[}
\newcommand{\rs}{\right]}

\begin{document}

\title{Treating Coulomb exchange
contributions in relativistic mean field
calculations: why and how}
\author{
  Nguyen Van Giai\email{nguyen@ipno.in2p3.fr} \\
  \it Institut de Physique Nucl\'eaire, Universit\'e Paris-Sud, F 91405 Orsay, France\\
  Haozhao Liang \\
  \it RIKEN Nishina Center, Wako 351-0198, Japan \\
  Huai-Qiang Gu and Wenhui Long \\
  \it School of Nuclear Science and Technology, Lanzhou University, Lanzhou 730000, China \\
  Jie Meng \\
  \it School of Physics, Peking University, Beijing 100871, China
}
\date{}
\maketitle

\begin{abstract}
The energy density functional (EDF) method is very widely used in nuclear physics, and among the various existing functionals
those based on the relativistic Hartree (RH) approximation are very popular because the exchange contributions (Fock terms)
are numerically rather onerous to calculate. Although it is possible to somehow 'mock up' the effects of meson-induced exchange terms by adjusting
the meson-nucleon couplings, the lack of Coulomb exchange contributions hampers the accuracy of predictions. In this note, we show that the
Coulomb exchange effects can be easily included with a good accuracy in a perturbative approach. Therefore, it would be desirable for future relativistic
EDF models to incorporate Coulomb exchange effects, at least to some order of perturbation.
\end{abstract}

\section{Introduction}

Our current microscopic understanding of the properties of atomic nuclei is based on two main approaches: using the nuclear shell model, on the one hand,
and the energy density Functional (EDF) method, on the other hand. The real start of application of EDF method to nuclear systems can be traced back to the early
1970s when Vautherin and Brink \cite{VB1972} revived the effective Skyrme interaction \cite{Sk1956} and proposed a local EDF for nuclear systems. Since then, a
vast number of articles dealing with a Skyrme-type EDF have appeared, and this trend will certainly continue for the years to come. The main reason for this
success is that the Skyrme EDF is a local functional, thus leading to local self-consistent mean fields. Even the Coulomb part of the Skyrme EDF was made local
through the Slater approximation for the exchange Coulomb contribution \cite{Slater}. The validity of the Slater approximation in Skyrme EDF calculations of
atomic nuclei has been discussed in the literature \cite{TQ1974}, and it was found to be accurate at the level of a few percent throughout the mass table with better results for
medium and heavy nuclei.

In the late 1980s a new type of EDF became also very popular, namely the so-called relativistic mean field (RMF) approach, which is essentially a relativistic
Hartree model with a no-sea assumption. The effective nucleon-nucleon interactions were mediated by meson exchanges with adjusted coupling strengths. From
Walecka's toy model of the early 1970s \cite{Walecka} it evolved into a sophisticated Hartree-type description of atomic nuclei \cite{Serot}.
Later on, it was extended into a relativistic Hartree-Bogoliubov (RHB) version to take into account the effects of nuclear pairing.
For the sake of numerical simplicity all exchange (Fock)
terms were dropped, their effects being hopefully taken care of by readjusting appropriately the meson-nucleon couplings.

The RMF and RHB approaches are generally quite successful \cite{Ring} and they are currently as widely applied as the non-relativistic Skyrme-type EDF approach.  However,
one aspect has been so far overlooked: the Coulomb exchange effects are absent from the RMF description, although it is well known that they are non-negligible,
especially in lighter systems (the asymptotic behaviour of the Coulomb mean field in a nucleus having $Z$ protons is $(Z-1)/r$, but it is $Z/r$ if only the Hartree
mean field is considered). In this contribution, we recall that a relativistic version of the Slater approximation is at hand, based on a local density approximation (LDA)
for the Coulomb exchange energy \cite{Gu2013}. This procedure leads to a set of RMF-type equations where the Coulomb part of the self-consistent mean field is a sum of a
direct (local) Coulomb potential, and an exchange Coulomb potential which is also local.

The relativistic Slater approximation for the Coulomb exchange field would be very useful for the newly developed covariant point-coupling models \cite{Zhao2010,Liang}
where meson-nucleon vertices are assumed of the contact type. Then, all nuclear and Coulomb mean fields would contain exchange contributions and still remain local.
In the rest of this short note, we will recall the main expressions which enable one to calculate the Coulomb exchange energy and potential using the
relativistic LDA (RLDA), and then we
comment on the first applications \cite{Gu2013} carried out recently with this method.

\section{Approximate forms of the relativistic Coulomb exchange energies and potentials}
In the non-relativistic mean field treatment of a finite nucleus having nucleonic densities $\rho_n(\mathbf{r})$ and $\rho_p(\mathbf{r})$ the Hartree-Fock (HF) single-particle
wave functions for protons are solutions of Schr\"odinger-type equations containing a Coulomb mean field $V_{\rm C}$. This Coulomb potential has a direct (Hartree)
component $V_{\rm Cdir}(\mathbf{r})$ which is local, and an exchange (Fock) non-local component $V_{\rm Cex}({\bf r, r'})$. The LDA consists in
approximating the Coulomb exchange energy density at point $\mathbf{r}$ of the inhomogeneous system by that of a homogeneous system having a proton density
value equal to $\rho_p(\mathbf{r})$. Since the single-particle wave functions in a homogeneous medium are plane waves, the Coulomb exchange energy per unit
volume can be expressed in the simple form \cite{Gu2013}:
\begin{equation}\label{eqA}
    e_{\rm Cex} = -\frac{3}{4}\lb\frac{3}{\pi}\rb^{1/3}e^2{\rho}_p^{4/3}\,.
\end{equation}
Under the LDA assumption, the Coulomb exchange energy of the finite system would then be
\begin{equation}\label{eqB}
    E_{\rm Cex}^{\rm LDA} = -\frac{3}{4}\lb\frac{3}{\pi}\rb^{1/3}e^2 \int d^3r\rho_p^{4/3}(\mathbf{r})\,.
\end{equation}
This is the so-called Slater approximation \cite{Slater} routinely employed in many self-consistent, non-relativistic mean field studies \cite{BHR2003}.
It leads to a local potential representing the exchange (Fock) Coulomb potential:
\begin{equation}\label{eqC}
    V_{\rm Cex}^{\rm LDA}(\mathbf{r}) =  \frac{\delta E_{\rm Cex}^{\rm LDA} }{\delta\rho_p(\mathbf{r})}
 =    -\lb\frac{3}{\pi}\rb^{1/3}e^2\rho_p^{1/3}(\mathbf{r})\,.
\end{equation}

In the relativistic case, the Coulomb energy is expressed again in terms of the proton density:
\begin{equation}\label{eqD}
    \rho_p(\mathbf{r}) =  \sum_i^p v_i^2   \bar{\psi}_i (\mathbf{r}) {\gamma}^0 {\psi}_i (\mathbf{r})
\end{equation}
and proton current:
\begin{equation}\label{eqE}
    \mathbf{j}_p(\mathbf{r}) = \sum_i^p v_i^2   \bar{\psi}_i (\mathbf{r}) \boldsymbol{\gamma} {\psi}_{i}(\mathbf{r})\,,
\end{equation}
where ($\gamma^0$, $\boldsymbol{\gamma}$) are the Dirac matrices, the $\psi_i$ are four-component spinors and the $v_i^2$ are occupation probabilities.
The summations run only over the proton states belonging to the positive energy spectrum (the no-sea approximation \cite{Walecka}). Then, the relativistic expressions
for the direct and exchange Coulomb energies are
\begin{align}\label{eqF}
    E^{\rm R}_{\rm Cdir}&=
    \frac{e^2}{2}\iint
    d^3r d^3r'\ls\frac{\rho_p(\mathbf{r})\rho_p(\mathbf{r}')}{|\mathbf{r}-\mathbf{r}'|}
    - \frac{\mathbf{j}_p(\mathbf{r})\cdot \mathbf{j}_p(\mathbf{r}')}{|\mathbf{r}-\mathbf{r}'|}\rs\,,\nonumber\\
\end{align}
\begin{align}\label{EqG}
    E^{\rm R}_{\rm Cex}&=-\frac{e^2}{2}\sum_{ij}^pv_i^2v_j^2\iint d^3r d^3r'
    \frac{\cos(|\varepsilon_i-\varepsilon_j||\mathbf{r}-\mathbf{r}'|)}{|\mathbf{r}-\mathbf{r}'|}\nonumber\\
    &\qquad\times\bar\psi_i(\mathbf{r})\gamma^\mu\psi_j(\mathbf{r})\bar\psi_j(\mathbf{r}')\gamma_\mu\psi_i(\mathbf{r}')\,,
\end{align}
where the $\epsilon_i$ are the single-particle energies.

In the relativistic homogeneous nuclear matter, the ${\psi}_{i}$ are plane wave solutions of the Dirac equation, and the time-like and space-like
components of $E^{\rm R}_{\rm Cex}$ per unit volume, $\bar{e}^{\rm R}_{\rm Cex}$ and $\bar{\bar{e}}^{\rm R}_{\rm Cex}$, can be related to the non-relativistic
energies  $e_{\rm Cex}$ of Eq.~(\ref{eqA}) through \cite{Engel}
\begin{align}\label{EqH}
   \bar{e}^{\rm R}_{\rm Cex} &= e_{\rm Cex} \bar\Phi(\beta)\,, \nonumber \\
   \bar{\bar{e}}^{\rm R}_{\rm Cex} &= e_{\rm Cex} \bar{\bar\Phi}(\beta)\,,
\end{align}
where
\begin{align}\label{newJ}
\beta=\frac{(3\pi^2n_p)^{1/3}}{M}\,,
\end{align}
$M$ being the proton mass, whereas $ \bar\Phi(\beta)$ and $\bar{\bar\Phi}(\beta)$ are analytical functions of $\beta$\cite{Gu2013}.
The relativistic corrections to the Coulomb exchange energy increase with the density, and they are substantial in atomic nuclei.
On the other hand, it is sufficient to evaluate them by expanding $\bar\Phi(\beta)$ and $\bar{\bar\Phi}(\beta)$
up to order $\beta^2$. To that order, the Coulomb exchange energy is
\begin{equation}\label{EqK}
    E^{\rm RLDA}_{\rm Cex}
    = -\frac{3}{4}\lb\frac{3}{\pi}\rb^{1/3}e^2\int d^3r\rho_p^{4/3}
        \ls 1-\frac{2}{3}\frac{(3\pi^2\rho_p)^{2/3}}{M^2}\rs\,,
\end{equation}
while the corresponding contribution to the single-particle potential for protons reads
\begin{equation}\label{EqL}
    V^{\rm RLDA}_{\rm Cex}(\mathbf{r})=
    -\left(\frac{3}{\pi}\right)^{1/3}e^2\rho_p^{1/3}(\mathbf{r})+\lb\frac{3\pi}{M^2}\rb e^2\rho_p(\mathbf{r})\,.
\end{equation}

\section{Application to nuclei}
Recently, the RLDA method for treating Coulomb exchange effects was checked by comparing it with the full treatment of Coulomb exchange in the relativistic Hartree-Fock-Bogoliubov calculations \cite{Gu2013}. Several isotopic chains were studied.

\begin{figure}
\includegraphics[width=8cm]{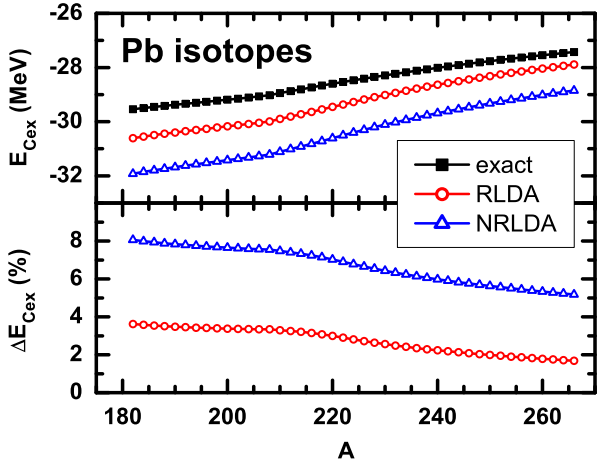}
\caption{\label{Fig1} Coulomb exchange energies in Pb isotopes calculated in RHFB approach with PKA1 model \cite{PKA1} (black squares). The LDA results calculated with
the same proton density distributions $\rho_p(r)$ but using the non-relativistic expression (\ref{eqB}) (resp., relativistic expression (\ref{EqK})) are shown by the NRLDA (resp., RLDA) curves.
The lower panel shows the percentage errors relative to the exact exchange energies.}
\end{figure}

As an example of application, we show in Fig.~\ref{Fig1} the Coulomb exchange energies $E_{\rm Cex}$ calculated for the chain of Pb isotopes. We have performed self-consistent
RHFB calculations of these nuclei, and then extracted the values of $E_{\rm Cex}$ corresponding to three choices: full RHFB (labeled ``exact'' in Fig.~\ref{Fig1}), non-relativistic LDA
(NRLDA, Eq.~(\ref{eqB})), and relativistic LDA (RLDA, Eq.~(\ref{EqK})). These results show that ignoring completely the Coulomb exchange energies introduces an error
of several tens of MeV which would have to be compensated artificially by readjusting the meson-nucleon couplings. On the other hand, the very simple RLDA expression
shown in
Eq.(10) for the Coulomb exchange energies in the $Z=82$ isotopes is accurate at the level of about $1$~MeV.
Thus, it would be an interesting improvement for models belonging to the RMF category to incorporate the Coulomb exchange effects by
the RLDA method
when adjusting the meson-nucleon couplings of the RMF Lagrangians. All the self-consistent potentials (including the exchange Coulomb potential)
entering the Dirac equations of the RMF would still
remain local, but a major part of the Coulomb exchange effects would be described by the local potential  $V^{\rm RLDA}_{\rm Cex}(\mathbf{r})$.

There are presently attempts to construct effective Lagrangians with point-coupling interaction vertices. Such models can describe nuclear systems within a relativistic
Hartree-Bogoliubov framework, and they can be generalized to RHFB form. The point-coupling assumption leads to local, self-consistent nuclear potentials and makes the Dirac equations easier to solve numerically.
For the Coulomb interaction, it is of course unreasonable to assume the point-coupling and a good strategy would be to handle fully the direct Coulomb interaction
while the exchange Coulomb effects would be described by the $V^{\rm RLDA}_{\rm Cex}(\mathbf{r})$ potential, similarly to what is done with Skyrme effective interactions.

Another observation that one can deduce from the comparison of the relativistic results of Ref.~\cite{Gu2013} with the study of non-relativistic Slater approximation carried out
by Titin-Schnaider and Quentin \cite{TQ1974} using a Skyrme-type EDF is that the RLDA method seems to give more accurate Coulomb exchange energies.
In Ref.~\cite{TQ1974} the $A \simeq 16$--$56$ region was explored, and it turned out that the non-relativistic Slater approximation overestimates the
Coulomb exchange energies by $5$--$7\%$. On the other hand, the relative error due to the RLDA is always less than $4\%$ (and could be of either sign) over a wide range of nuclides
from $A=40$ to $A=266$.

Thus, one has now a simple and efficient way to incorporate into RMF-type calculations, or more generally into RHFB models with 
point-coupling vertices,
most of the effects due to Coulomb exchange interactions among protons inside a nucleus.


\begin{ack}
This work was partly supported by
the Fundamental Research Funds for Central Universities under Contracts No. lzujbky-2011-15 and No. lzujbky-2012-k07,
the Major State 973 Program 2013CB834400,
the National Natural Science Foundation of China under Grants No. 10975008, No. 11075066, No. 11105006, and No. 11175002,
the Research Fund for the Doctoral Program of Higher Education under Grant No. 20110001110087,
the Grant-in-Aid for JSPS Fellows under Grant No. 24-02201,
and the Program for New Century Excellent Talents in University of China under Grant No. NCET-10-0466.
\end{ack}


\end{document}